\documentclass[twocolumn,prc,preprintnumbers,nofootinbib,superscriptaddress,showpacs,a4]{revtex4-2}
\usepackage{epsfig,graphicx,float,pst-all}
\usepackage{mathrsfs}
\usepackage[utf8]{inputenc}
\usepackage{rotating}
\usepackage{url}
\usepackage{esvect}
\usepackage{subfigure}
\usepackage{multirow}
\usepackage{amsmath}
\usepackage{amssymb}
\usepackage{amsfonts}
\usepackage{bm}
\usepackage{xcolor}
\usepackage{natbib}
\usepackage{float}
\usepackage{hyperref}
\usepackage[warn]{textcomp}
\usepackage{gensymb}
\usepackage{footnote}
\usepackage{siunitx}
\usepackage{tabularx}
\usepackage{physics}
\usepackage{comment}
\usepackage{bm}
\usepackage{orcidlink}
\usepackage[normalem]{ulem}

\newcolumntype{C}[1]{>{\centering\arraybackslash}X}

\begin{document}
	\title{Simultaneous calculation of elastic scattering, transfer, breakup, and other direct cross sections for $d$+$^{197}$Au reaction} 
	\author{H. M. Maridi\orcidlink{0000-0002-2210-9897}}
	\email[Corresponding author: 
	]{hasan.maridi@manchester.ac.uk}
	\affiliation{Department of Physics and Astronomy, The University of Manchester, Manchester M13 9PL, UK}
	\author{D. K. Sharp\orcidlink{0000-0003-1167-1603}}
    \affiliation{Department of Physics and Astronomy, The University of Manchester, Manchester M13 9PL, UK}
	\author{J. Lubian\orcidlink{0000-0001-7735-1835}}
	\affiliation{Instituto de F\'{\i}sica, Universidade Federal Fluminense, Avenida Litoranea s/n, Gragoat\'{a}, Niter\'{o}i, Rio de Janeiro 24210-340, Brazil}
	\date{\today}

	\begin{abstract}
		Simultaneous analyses are performed for cross section of elastic scattering, Coulomb breakup, transfer, and other direct yields for the $d$+$^{197}$Au system at all available energies. The data are reproduced well by the optical model that is based on parts related to the Coulomb and nuclear contributions of the direct cross sections. This method of calculation can be successfully applied to the reactions of deuteron with heavy targets.
	\end{abstract}
		

	\maketitle

\section{\label{sec:int} Introduction}
Reactions induced by deuterons attracted the attention of nuclear physicists from the very beginning of the appearance of this research area. The main reason for that is the simplicity of the deuteron structure, which is concerned with the number of nucleons. This also simplifies the study of the reaction mechanism due to the limited number of projectile particles participating. The deuteron is formed by one proton and one neutron bound by the action of the attractive nuclear interaction only with a low breakup threshold (2.2245 MeV). It has only one bound state, the ground state. For this reason, in the interaction with different targets, several reactions can occur, like elastic scattering, inelastic excitation of the target, neutron striping, or pick-up, breakup, and fusion of the whole projectile (known as complete fusion) or of the proton or the neutron after its breakup (known as incomplete fusion). The deuteron is an excellent probe to access the single-particle structure of the target in the neutron transfer reactions. Many theoretical and experiential papers have been reported in the literature about reactions emitting protons and neutrons from the interaction of deuteron with different targets. Some reports on this topic can be found in Refs.~\cite{BaT76,AIK87,BRT84}. Some recent studies of deuteron-induced reactions were also reported~\cite{CLN23,RTN16,KLN18,AWK17,ARH23,Des22,PPC17,CNC17}.
Different theoretical models can be used to account for all these reaction channels. Among them, the one-channel or optical model calculation is usually used to determine the reaction cross section and to study the energy dependence of the optical potential~\cite{CDG15}. The coupled channel method is used to determine the cross section for the inelastic excitations. The coupled reaction method is used to study rearrangement reactions. Finally, the continuum discretized coupled channel method to account for the elastic breakup.
Besides all these theoretical methods, an energy-dependent optical model potential under some physical constraints can be used to simultaneously reproduce the elastic scattering angular distribution, transfer, and reaction cross-section.

In this work, we use this last method to describe data for the $d$+$^{197}$Au reaction at a wide range of energy.


\section{\label{sec:total} Total optical potential}
There are many deuteron global optical model potentials (OPs), namely, Daehnick \cite{Dae80}, Bojowald \cite{Boj88}, An \cite{An06}, and Han \cite{Han06}. 
They are expressed within Woods-Saxon (WS) shape in the $d$-$T$ coordinate, $r$, as
\begin{eqnarray}\label{eq:op}
	U(r) = -V_{\textrm{r}}f_{\textrm{r}}(r)-iW_{\textrm{v}}f_{\textrm{v}}(r)+i 4a_\textrm{s} W_{\textrm{s}}\frac{d}{dr}f_{\textrm{s}}(r) \nonumber \\
	+\lambda_{\pi}^2 V_{\textrm{so}}\frac{1}{r}\frac{d}{dr}f_{\textrm{so}}(r) {\bf \sigma.\textit{l}}+ V_{\textrm{C}}(r),
\end{eqnarray}

where $V_{\textrm{r}}$, $W_{\textrm{v}}$, $W_{\textrm{s}}$, and $V_{\textrm{so}}$ are depths of the real, volume imaginary, surface imaginary, and spin-orbit potentials, respectively.
$f_{\textrm{i}}(r)=\frac{1}{1+\exp[(r-R_i)/a_i]}$ 
with $R_i=r_i A_{T}^{1/3}$ 
where $A_\textrm{T}$ is the target mass number.
The Coulomb potential is
\begin{equation}
	V_{\textrm{C}}(r)=
	\begin{cases}
		\frac{\displaystyle Z_{\textrm{P}}Z_{\textrm{T}}e^2}{\displaystyle r},
		& (r\geqslant R_{\textrm{C}})\\
		\frac{\displaystyle Z_{\textrm{P}}Z_{\textrm{T}}e^2}{\displaystyle
			2R_{\textrm{C}}}\left(3-\frac{\displaystyle
			r^2}{\displaystyle R_{\textrm{C}}^2}\right)  & (r\leqslant R_{\textrm{C}}).
	\end{cases}
\end{equation}
where $Z_{\textrm{T}}$ and $Z_{\textrm{P}}$ are the charge numbers of the target and the projectile nuclei, respectively, and $R_\textrm{C}=1.3 A_{T}^{1/3}$ is the radius of the Coulomb potential.
Generally, the imaginary OP consists of volume and surface absorption components, 
the volume part which is often arranged to simulate the ingoing-wave boundary condition to model flux loss due to fusion and the surface part that accounts for flux loss due to non-elastic direct reaction channels \cite{Kee09}.

In this work, to stimulate the long-range interactions, instead of the surface imaginary OP in the global OP (\ref{eq:op}), the surface potential will consist of three components: (i) the Coulomb dynamical polarization potential (CDPP) which represents the Coulomb breakup; (ii) the imaginary nuclear dynamical polarization potential (NDPP) for the nuclear transfer; and (iii) a surface imaginary potential for the nuclear breakup and other missing direct channels.

The CDPP can be obtained by solving the formalism for the scattering of deuteron (as a proton core and valence neutron) from a heavy target and can be given  
as \cite{Mar21,Mar24}
\begin{eqnarray}
	\label{eq:CDPP}
	\delta V_{\textrm{C}} (r)&=& \varepsilon_0 \left[ \frac{QG_{0}F_{0}+Q^2 G_{0}F_{0}G_{0}'F_{0}' +Q^2 F_{0}^2 F_{0}'^2}{F_{0}^4 + G_{0}^2 F_{0}^2}-1 \right] \nonumber \\
	\delta W_{\textrm{C}} (r)&=& \varepsilon_0 \left[ \frac{Q^2 F_{0}F_{0}' -Q F_{0}^2}{F_{0}^4 + G_{0}^2 F_{0}^2} \right]
\end{eqnarray}
where $\delta V_\textrm{C} (r)$ and $\delta W_\textrm{C} (r)$ are the real and imaginary parts of the CDPP, respectively, $F_0$ and $G_0$ the regular and irregular Coulomb functions in $\rho= k(r)r$ and $\eta= \frac{m_p^2}{\mu} \frac{Z_\textrm{P} Z_\textrm{T} e^2}{\hbar^2 k(r)}$. $Q(r) =\frac{\mu}{m_{p}} \frac{k(r)} {\kappa_0}$ with $k(r) \approx {\sqrt{\frac{2 m_{p}^2}{\mu {\hbar}^2}(V_\textrm{C} ({ r})+\varepsilon_0^*)}}$ and $\kappa_0 = \sqrt{\frac{-2\mu_p\varepsilon_0}{\hbar^2}}$, where $\mu$ is the deuteron reduced mass, $m_{p}$ the proton mass, and $\varepsilon_0$ the deuteron separation energy.

The NDPP utilizes a surface WS-type potential, usually with large radius and/or diffuseness parameters. 
Then, the $(d,p)$ transfer can be represented by the imaginary long-range NDPP \cite{Mar24}
\begin{eqnarray}
	\label{eq:NDPP}
	\delta W_{d,p} (r) =
	4a_\textrm{L} W_{\textrm{L}}\frac{d}{dr}f_{\textrm{L}}(r)
\end{eqnarray}
with the fixed geometry parameters extracted from the semiclassical theory \cite{Bon02} with a strong absorption radius $R_\textrm{L}=1.4 (A_\textrm{P}^{1/3}+A_\textrm{T}^{1/3})$ and a diffuseness linked to the separation energy $\varepsilon_0$ as $a_\textrm{L}=\hbar/\sqrt{-8 \mu \varepsilon_0}$. For deuteron: $a_\textrm{L}=2.167$ fm since $\varepsilon_0=-2.225$ MeV. The depth, $W_\textrm{L}$, is varied to fit the transfer cross section data.

Now, the total deuteron-target optical potential becomes
\begin{eqnarray}\label{eq:op2}
	U(r) = V_{\textrm{C}}(r)+\delta V_{\textrm{C}} (r)-V_{\textrm{r}}f_{\textrm{r}}(r)-iW_{\textrm{v}}f_{\textrm{v}}(r) \nonumber \\	
	+i 4a_\textrm{s} W_{\textrm{s}}\frac{d}{dr}f_{\textrm{s}}(r)
	+i 4a_\textrm{L} W_{\textrm{L}}\frac{d}{dr}f_{\textrm{L}}(r) \nonumber \\
	+i\delta W_{\textrm{C}} (r)
	+\lambda_{\pi}^2 V_{\textrm{so}}\frac{1}{r}\frac{d}{dr}f_{\textrm{so}}(r) {\bf \sigma.\textit{l}}.
\end{eqnarray}
To fit the data, we keep the original global parameters \cite{An06} and vary only three parameters: (i) the imaginary NDPP depth ($W_\textrm{L}$) to fit the transfer cross section data; (ii) the depth of the real OP ($V_\textrm{r}$); and (iii) the imaginary surface depth ($W_\textrm{s}$) to fit the elastic scattering data. 	
Here, we would like to clarify that the new surface imaginary potential in (\ref{eq:op2}) represents the remaining direct reaction contributions such as nuclear breakup. It differs from the earlier surface OP in (\ref{eq:op}) that represents all direct contributions. So we have chosen to fix the geometry parameters to $r_\textrm{s}=1.4$ fm and $a_\textrm{s}=0.76$ fm instead of the geometry used in the surface potential of Ref.\cite{An06}, $r_\textrm{s}=1.36$ fm and $a_\textrm{s}=0.89$ fm. We note that our model allows us to add many surface OPs with different geometries to represent the different direct channels. In contrast, in global and phenomenological OP, there is only one surface OP for all direct reaction channels.

The partial reaction cross sections can be calculated from the imaginary potentials as
\begin{eqnarray}
	\label{eq:sigr3}
	\sigma_{i} =  \sum_{\ell} \sigma_{i;\ell} =- \frac{2}{\hbar \upsilon} \frac{4\pi}{k^2} \sum_{{\ell}=1}^{\infty} (2{\ell}+1)\int dR |\chi_{\ell} (r)|^2 W_{i}(r) \nonumber \\
\end{eqnarray}
where $\chi_{\ell} (r)$ is the partial-wave radial functions, $\upsilon$ is the asymptotic relative velocity. Then we have  $\sigma_{vol}$, $\sigma_{sur}$, $\sigma_{\mathrm{d,p}}$, $\sigma^{C}_{\mathrm{bu}}$ which are related to the imaginary potentials $W_{\textrm{v}}$, $W_{\textrm{s}}$, $\delta W_{d,p}$, $\delta W_{\textrm{C}}$, respectively.
\begin{figure*}[!ht]
	\centering
	\includegraphics[width=0.3\textwidth,clip=]{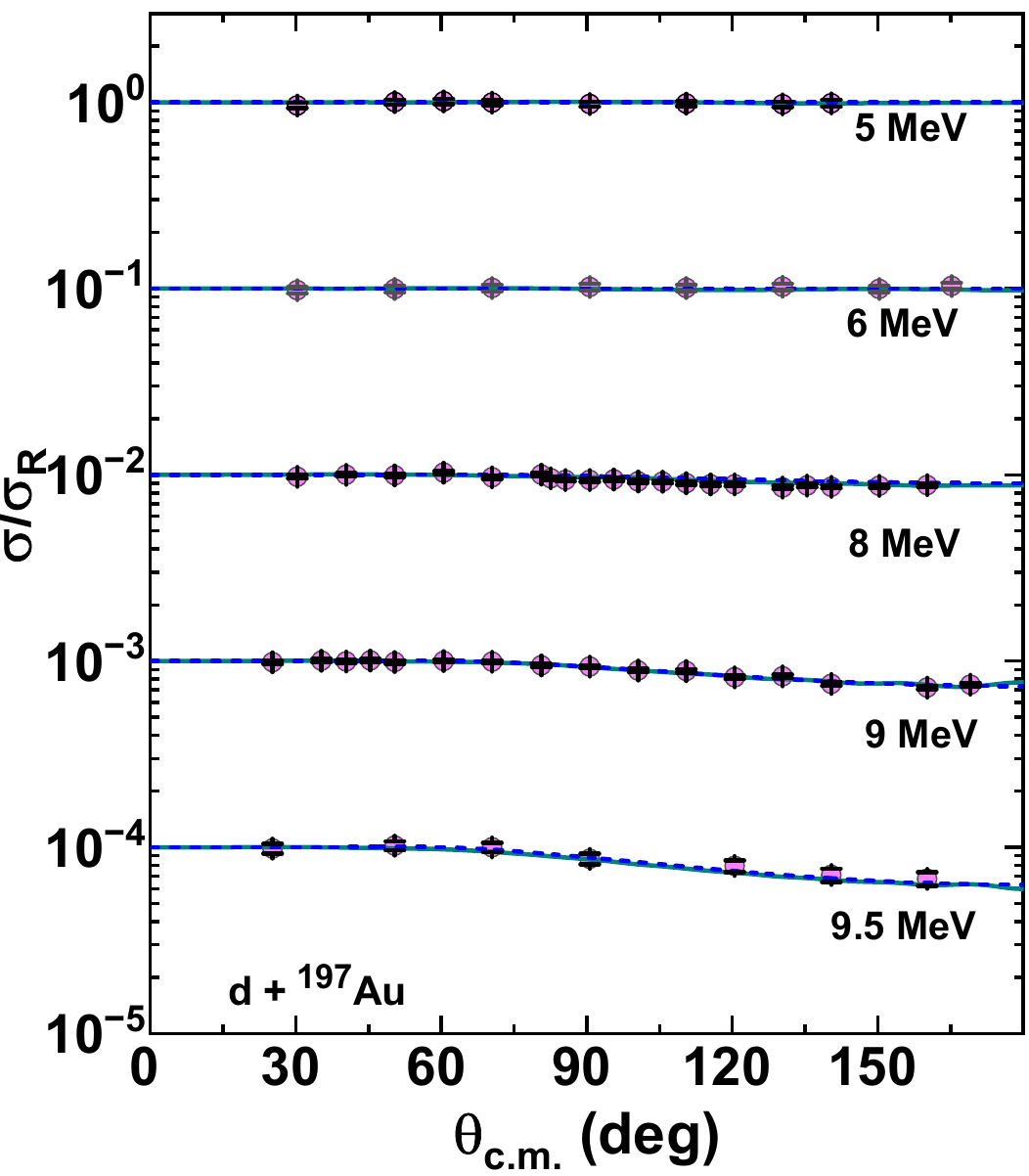}
	\includegraphics[width=0.3\textwidth,clip=]{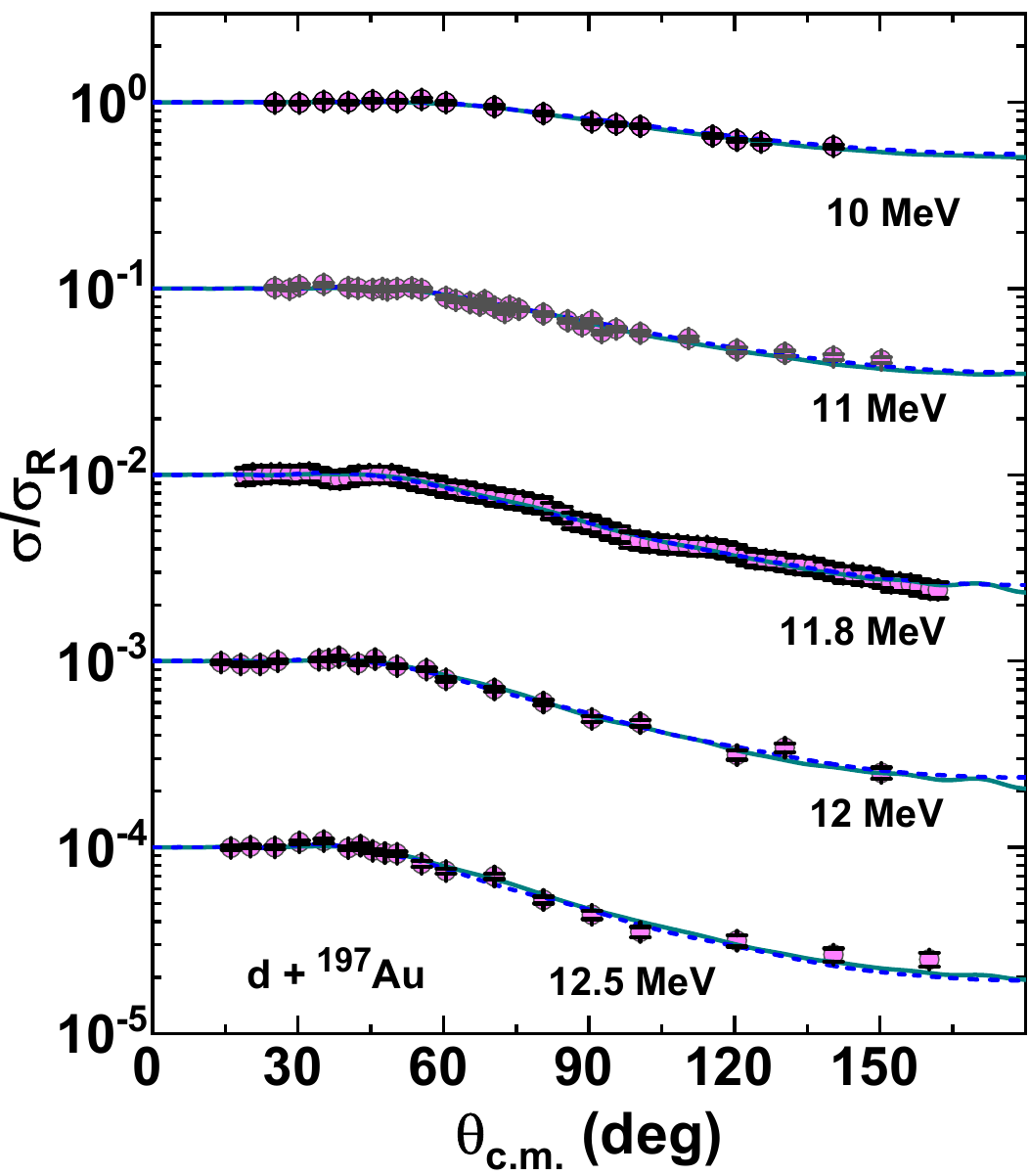}
	\includegraphics[width=0.3\textwidth,clip=]{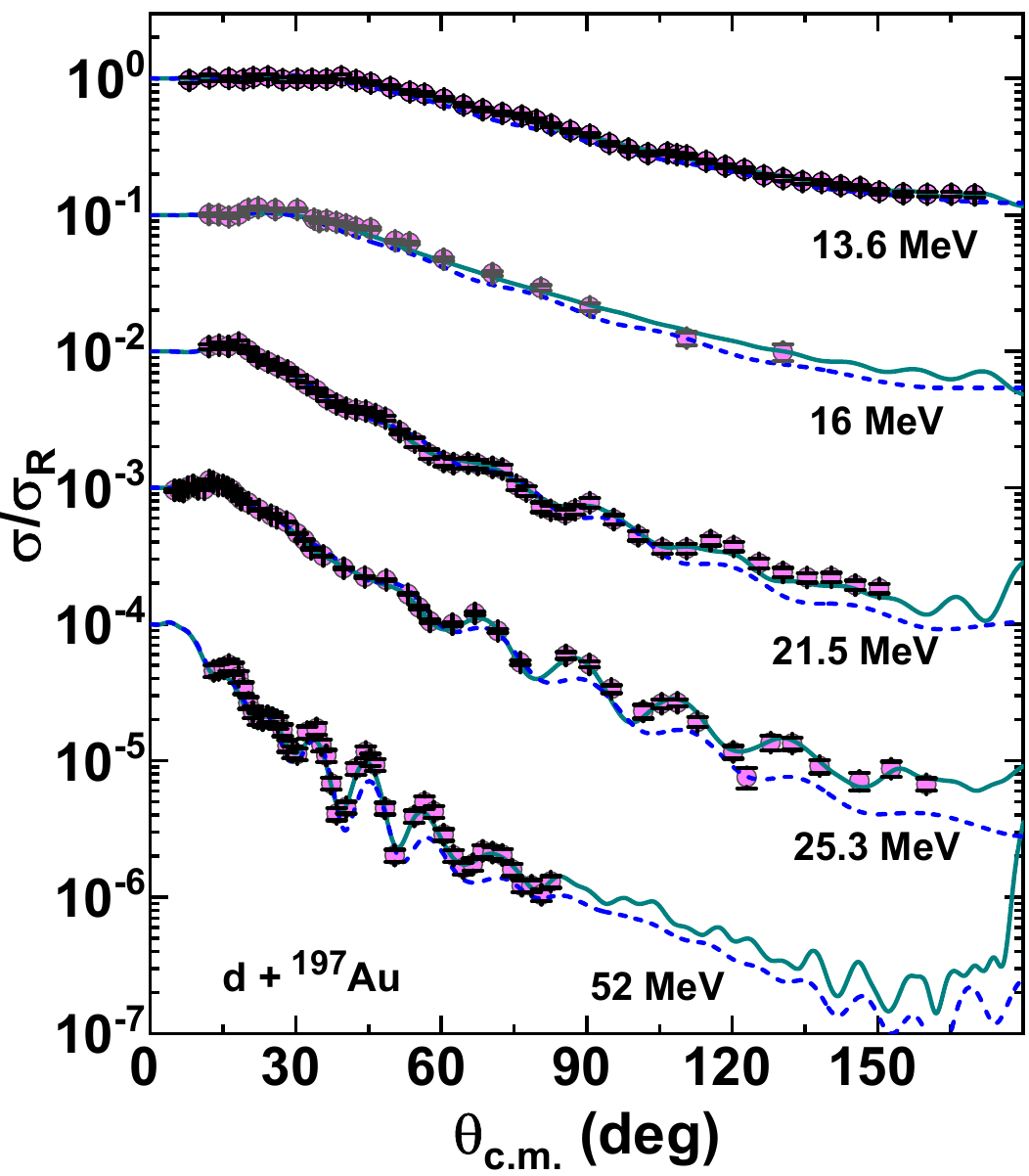}
	\caption{\label{fig:de} The calculated $d$+$^{197}$Au elastic scattering angular distributions compared with the data \cite{Giu23,Igo61,Kor68,Ynt59,Dji64,Hin68}. The solid lines refer to our calculations [Eq. (\ref{eq:op2})] and the dashed lines represents the deuteron-global OP \cite{An06}.}
\end{figure*}

\section{\label{sec:app} Results and discussion}

\begin{table}
	\centering
	\caption{\label{tab:cs} Calculated cross sections for $d$-$^{197}$Au system. $E_{lab}$, $V_{\textrm{r}}$, $W_{\textrm{s}}$, $W_{\textrm{L}}$ are in MeV and  $\sigma$'s in mb.}
	\begin{tabular}{lllllllll}
		\hline\noalign{\smallskip}
		$E_{lab}$ &$V_{\textrm{r}}$  &  $W_{\textrm{s}}$&$W_{\textrm{L}}$ & $\sigma_{vol}$&$\sigma_{sur}$ &$\sigma_{\mathrm{d,p}}$&$\sigma^{\textrm{C}}_{\mathrm{bu}}$  &$\sigma_{R}$   \\
		\hline\noalign{\smallskip}
		5	 &100	 &15	 &0.01  &0	     &0.1	 &0.2	 &0.1	 &0.4 \\
		6	 &100	 &15	 &0.02  &0	     &1.3	 &2.4	 &1.1	 &4.9 \\
		8	 &100	 &15	 &0.033 &0.2	 &61.3	 &17.8	 &16.6	 &95.9 \\
		9	 &75.8	 &6.65	 &0.055 &2.4	 &111	 &58.2	 &38.0	 &209 \\
		9.5	 &80	 &6.6	 &0.07  &4.4	 &172	 &87.1   &50.7	 &314 \\
		10	 &90	 &6.55	 &0.088 &8.3	 &252	 &124	 &66.2	 &451 \\
		11	 &80	 &6.5	 &0.111 &15.4	 &431	 &184	 &91.0	 &721 \\
		11.8 &60	 &4.5	 &0.114 &33.5	 &463	 &216	 &122	 &835 \\
		12	 &64.7	 &4.5	 &0.115 &38.4	 &496	 &222	 &131	 &888 \\
		12.6 &44.1	 &4.5	 &0.122 &38.0	 &556	 &246	 &129	 &968 \\
		13.6 &26.4	 &3.55	 &0.113 &57.0	 &629	 &251	 &151	 &1088 \\
		16	 &20	 &5	     &0.105 &62.3	 &1026   &241	 &153	 &1482 \\
		21.6 &73	 &7.4	 &0.055 &123	 &1686   &123	 &199	 &2131 \\
		25.3 &98	 &7.9	 &0.035 &165	 &1909   &76.5	 &221	 &2371 \\
		52	 &90	 &9.65	 &0.014 &300	 &2447   &25.1	 &192	 &2965 \\
		\hline\noalign{\smallskip}	 	\end{tabular}
\end{table}
\begin{figure*}
	\centering
	\includegraphics[width=0.39\textwidth,clip=]{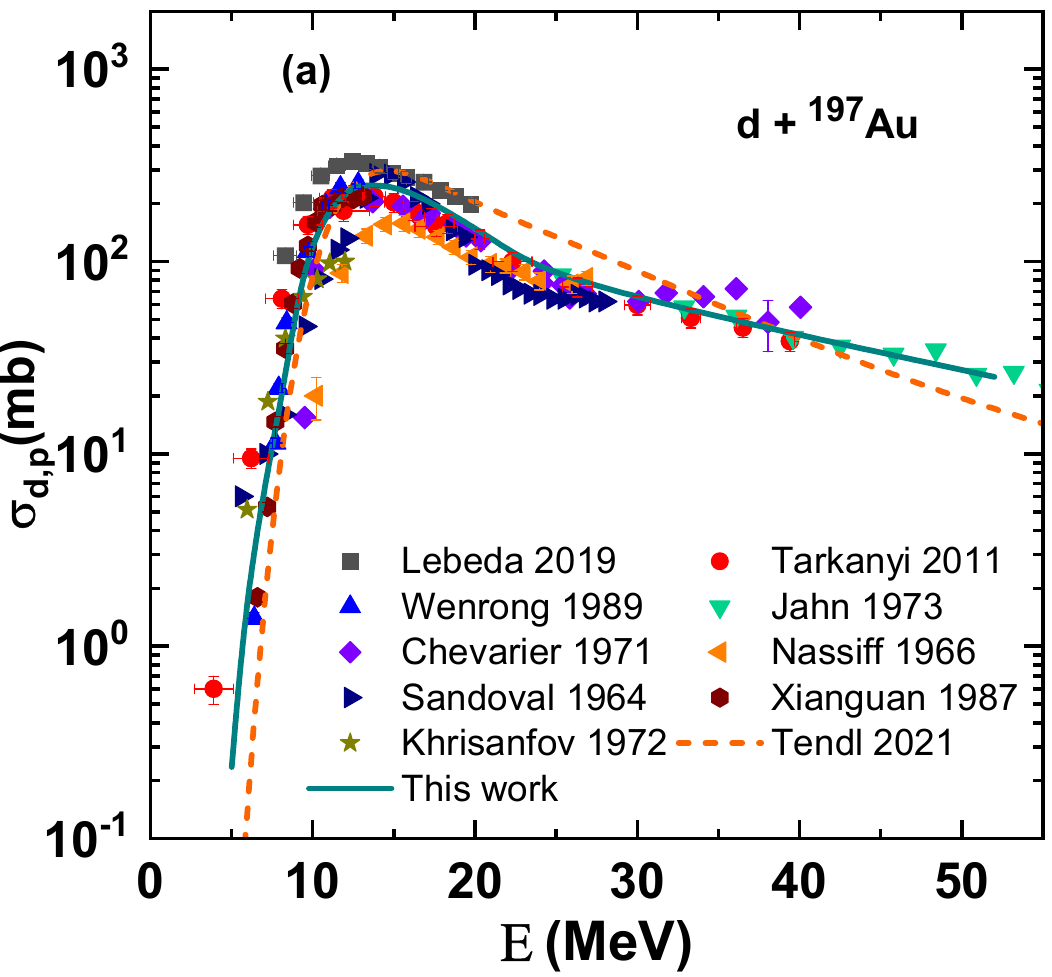}
	\includegraphics[width=0.39\textwidth,clip=]{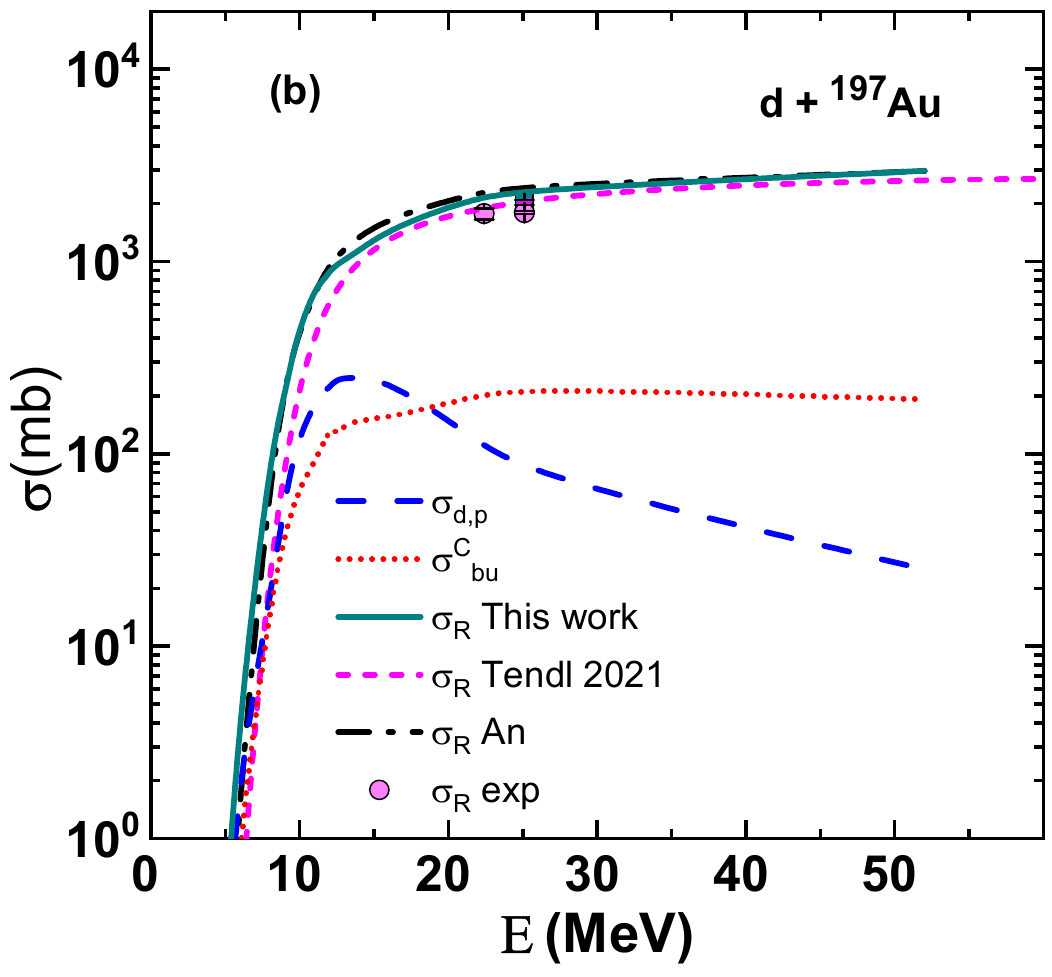}		
	\caption{\label{fig:sig} (a) The $^{197}$Au(d,p)$^{198}$Au transfer cross sections calculated using Eqs. (\ref{eq:sigr3}) and (\ref{eq:NDPP}) in comparison with the TALYS prediction from Tendl-2021 \cite{Tendl21} and experimental data from EXFOR database \cite{Exfor}. (b) The calculated transfer, Coulomb breakup, and total reaction cross sections 
		from the present work 
		in comparison with the TALYS prediction from Tendl-2021 \cite{Tendl21} and global OP \cite{An06} calculations and experimental data \cite{May65,Wil62}.}
\end{figure*}
Now we will apply our model to analyse $d$+$^{197}$Au reaction. Recently, the differential cross sections of $d$+$^{197}$Au elastic scattering were measured at ten incident energies between 5 and 16 MeV \cite{Giu23}. We take these data in addition to the other available elastic scattering data, namely, 11.8 \cite{Igo61}, 13.6 \cite{Kor68}, 21.6 \cite{Ynt59}, 25.3 \cite{Dji64} and 52 MeV \cite{Hin68}.
The results of the calculation are listed in Table \ref{tab:cs} and Figs. \ref{fig:de} and \ref{fig:sig}.
It is clear that our potential fits the elastic scattering data with an agreement better than the global OP, especially at energies above the Coulomb barrier, $V_{B} \sim 10.5$ MeV, as shown in Fig. \ref{fig:de}.
In addition, the ($d$,$p$) cross sections are reproduced well, as shown in Fig. \ref{fig:sig} (a), and the calculated total reaction cross sections are in agreement with the available data and the TALYS prediction adopted from the TENDL library. \cite{Tendl21}.
The fitted values of \( V_r(E) \), \( W_d(E) \), and \( W_L(E) \) can be expressed as
\begin{equation}
	f(E) = a_0 + \frac{a}{w^2 + 4(E - E_0)^2},
\end{equation}
where \( a_0 \), \( a \), \( E_0 \), and \( w \) are given, in MeV, as follows:  
\( a_0 = 99.64,\ 5.31,\ 0.00514 \);  
\( a = -2290.59,\ 152.78,\ 9.95 \);  
\( E_0 = 14.95,\ 6.14,\ 13.80 \);  
and \( w = 4.88,\ 3.68,\ 8.96 \),  
for \( V_r(E) \), \( W_d(E) \), and \( W_L(E) \), respectively.


\section{\label{sec:summary} Summary and conclusions}
In summary, we present simultaneous calculations for the elastic scattering, Coulomb breakup, neutron transfer, and other direct nuclear channels of the $d$+$^{197}$Au system within the optical-model framework. The optical potential used is based on the global deuteron OP and modified to include the polarization potentials that represent the Coulomb breakup and neuron transfer. The available data of the elastic scattering, ($d$,$p$) transfer cross section, and reaction cross section are reproduced well. The few fitted parameters are found to have a systematic behaviour with energy. 
The present model can analyse the deuteron scattering data over a range of near-barrier energies with a few parameters. 

\bigskip
\section*{Acknowledgments}
This work was funded by the Council for At-Risk Academics (Cara) within the Cara Fellowship Programme \& the British Academy within the British Academy/Cara/Leverhulme Researchers at Risk Research Support Grants Programme under grant number LTRSF/100141. J.L. thanks CNPq, FAPERJ, and INCT-FNA (Instituto Nacional de Ci\^{e}ncia e Tecnologia - F\'{i}sica Nuclear e Aplica\c{c}\~{o}es) research project 464898/2014-5 for the partial financial support.

\end{document}